1 **Nanoscale metallic iron for environmental remediation: prospects and limitations**

2 Chicgoua Noubactep[(a,d),*], Sabine Caré[(b)], Richard Crane[(c)]

3 [(a)] Angewandte Geologie, Universität Göttingen, Goldschmidtstraße 3, D - 37077 Göttingen, Germany.

4 [(b)] Université Paris-Est, Laboratoire Navier, (ENPC/IFSTTAR/CNRS), 2 allée Kepler, 77420 Champs sur Marne,

5 France.

6 [(c)] Interface Analysis Centre, University of Bristol, 121 St. Michael's Hill, Bristol, BS2 8BS, UK.

7 [(d)] Kultur und Nachhaltige Entwicklung CDD e.V., Postfach 1502, D - 37005 Göttingen, Germany.

8 Correspond author; e-mail: cnoubac@gwdg.de; Tel. +49 551 39 3191, Fax. +49 551 399379

10 **Abstract**

11 The amendment of the subsurface with nanoscale metallic iron particles (nano-$Fe^0$) has been

12 discussed in the literature as an efficient in-situ technology for groundwater remediation.

13 However, the introduction of this technology was controversial and its efficiency has never

14 been univocally established. This unsatisfying situation has motivated this communication

15 whose objective was a comprehensive discussion of the intrinsic reactivity of nano-$Fe^0$ based

16 on the state-of-the art knowledge on the mechanism of contaminant removal by $Fe^0$ and a

17 mathematical modelling. It is showed that due to limitations of the mass transfer of nano-$Fe^0$

18 to contaminants, available concepts can not explain the success of nano-$Fe^0$ injection for in-

19 situ groundwater remediation. It is recommended to test the possibility of introducing nano-

20 $Fe^0$ to initiate the formation of roll-fronts which propagation would induce the reductive

21 transformation of both dissolved and adsorbed contaminants. Within a roll-front, $Fe^{II}$ from

22 nano-$Fe^0$ is the reducing agent for contaminants. $Fe^{II}$ is recycled by biotic or abiotic $Fe^{III}$

23 reduction. While the roll-front concept could explain the success of already implemented

24 reaction zones, more research is needed for a science-based recommendation of nano-$Fe^0$ for

25 subsurface treatment by roll-fronts.

26 **Keywords**: Environmental remediation, Material reactivity, Nanoscale iron, Roll-front,

27 Zerovalent iron.

## 1    Introduction

The development of new methods and materials for environmental remediation is a real challenge for the scientific community. Such technologies will only be adopted by industry if they can exhibit marked improvements in efficiency, affordability or eco-compatibility compared to conventional techniques. The use of metallic iron ($Fe^0$) in subsurface reactive permeable barriers has been identified as such a technology [1-4]. Since this discovery almost 20 years ago, extensive research of $Fe^0/H_2O$ system has been performed in an attempt to understand the controlling mechanisms behind the remediation of redox-amenable contaminant species using $Fe^0$-based materials [5-14]. Two different tools are commonly used to optimise the efficiency of $Fe^0$ for aqueous contaminant removal: (i) reducing the particle size of $Fe^0$ down to the nanoscale (nano-$Fe^0$) [15,16], and (ii) using bimetallic systems [17,18]. In recent years there has been considerable interest into combining the two methods [19-23].

Since the original proof of concept study into the application of nano-$Fe^0$ for water treatment at Lehigh University, USA [15], research within this field has boomed. On April 23th 2011, a search at "**Science Direct"** using key words "*nanoscale*" and "*zerovalent iron*" yielded 208 peer-reviewed articles in 6 selected journals (Table 1). The same search at "**Environmental Science and Technology**" resulted in 157 articles. According to Table 1, 59 articles have already been published in the first quarter of 2011 in the 7 selected journals. This clearly demonstrates the interest within academia for this technology.

In recent years, several review articles and critical views on nano-$Fe^0$ for environmental remediation have been published [20, 24-40]. However, the original discussion on the suitability of nano-$Fe^0$ for in-situ field applications [25] has not been satisfactorily addressed [28,36]. Moreover, a recent comparison between field applications of $Fe^0$ of different particle sizes (nm, μm, and mm) for field applications has clearly demonstrated the superiority of mm-$Fe^0$ (average efficacy 97 %) [11]. The decreasing order of reactivity was mm-$Fe^0$ (97 %) >

μm-Fe[0] (91 %) > nano-Fe[0] (65 %). Expectably, the lower efficiency of nano-Fe[0] is due their high reactivity [25,34]. Therefore, the question arises on the fundamental necessity to further increase the reactivity of nano-Fe[0] by using a noble metal combination.

## 1.1    The problem

Nano-Fe[0] technology for environmental remediation was introduced as an alternative to the conventional Fe[0] walls mostly for inaccessible aquifers [27,41]. The very small particle size of nano-Fe[0] (1–100 nm) would allow the material to penetrate deep into soil networks [11,12,20,39,40,42].

Due to the exponential relationship between specific surface area (SSA) and radius (R = d/2) of a perfectly spherical object (SSA = $4\pi R^2$), as a rule, a decrease in Fe[0] particle size increases the surface area per gram by up to 3 orders of magnitude [22,29]. In other words, the inverse relationship between Fe[0] particle size and reactivity is due to a greater density of reactive sites on the particle surface at smaller scale. The following three claims have been made with regard to the use of nano-Fe[0] for aqueous contaminant removal (ref. [12] and ref. therein): (i) some aqueous contaminant species that have been proven as unsuccessful for remediation using μm-Fe[0] and mm-Fe[0] can be effectively removed using nano-Fe[0], (ii) nano-Fe[0] can be used for more rapid degradation of contaminants, and (iii) the formation of some undesirable by-products during remediation using μm-Fe[0] and mm-Fe[0] can be avoided by using nano-Fe[0]. Such processes whilst correct are all linked to the greater reactivity nano-Fe[0] possesses due to its size (reactive surface area). When performed in conditions without a large nano-Fe[0] stoichiometric excess, e.g. a system analogous to the environment, it may prove that such claims will be unfounded [21,34,36,43,44].

An undisputed drawback with regards to the use of nano-Fe[0] for environmental applications is their strong tendency to aggregate and adhere to solid surfaces [11,12,20,27,30,39,40,43]. Karn et al. [20] listed some parameters that influence nano-Fe[0] adsorption onto soil and aquifer materials: (i) the surface chemistry of soil and Fe[0] particles, (ii) the groundwater

80    chemistry (e.g., ionic strength, pH and presence of natural organic matter), and (iii) the

81    hydrodynamic conditions (pore size, porosity, flow velocity and degree of mixing or

82    turbulence). Several methods have been developed for the stabilization of nano-$Fe^0$ particles

83    over the past decade and proven efficient to sustain the reactivity of nano-$Fe^0$ [11,12,20,43].

84    One factor that has been overlooked, however, is the impact volumetric expansion has on the

85    mobility of (i) residual $Fe^0$, (ii) primary corrosion products ($Fe^{II}$ and $H_2$) and contaminants.

86    The volume of any corrosion product (Fe hydroxide or oxide) is higher than that of the

87    original metal ($Fe^0$). The ratio between the volume of expansive corrosion product and the

88    volume of iron consumed in the corrosion process is called "rust expansion coefficient" ($\eta$)

89    [45-47]. Volumetric corrosion products are likely to: (i) contribute to porosity loss, (ii) impact

90    the retention of contaminants and transformation products, and (iii) increase the particle

91    agglomeration.

92    Another area of heightened research is with regard to the determining the toxicity of nano-

93    $Fe^0$, with mixed results reported [12,48]. For example, Barnes et al. [49] reported minimal

94    change to the structure of a river water community due to the addition of nano-$Fe^0$, while

95    Diao and Yao [50] reported nano-$Fe^0$ particles as highly cytotoxic towards both gram-positive

96    and gram-negative bacteria species.

97    While taking into account all known influencing parameters, the following seven features

98    have to be systematically studied in order to optimise the general applicability of this

99    technique [12,20,51]: (i) mobility changes due to nano-$Fe^0$ volumetric expansion during

100    corrosion, (ii) the bioavailability of $Fe^0$ and corrosion products ($Fe^{II}/Fe^{III}$ species, $H/H_2$), (iii)

101    the ecotoxicity of $Fe^0$ and its corrosion products, (iv) the bioaccumulation of $Fe^0$ and its

102    corrosion products, (v) the translocation potential of nano-$Fe^0$, (vi) the long-term reactivity of

103    nano-$Fe^0$ particles, and (vii) the speciation, persistence and fate of contaminants and their

104    transformation products. A major contributing factor to the latter point is that little is known

105 (compared to permeable reactive barrier technologies) about the extent contaminants are

106 removed via size exclusion using nano-Fe$^0$.

107 Only when all seven "operational drivers" have been determined can the global community

108 have full faith in the technology.

109 **1.2    Objectives of the study**

110 The present communication is focused on the "field persistence" or reactive "life span" of

111 nano-Fe$^0$ particles. For in-situ applications a keen understanding of nano-Fe$^0$ reactive fate is

112 essential for effective and prudent site clean-up. The knowledge of which is likely to largely

113 underpin decisions as to the (i) the choice of material selected, (ii) the mechanism of

114 application and, (iii) the strategy (if any) for repeated treatments.

115 In the current work a multidisciplinary approach is used to analyse the relationship between

116 nano-Fe$^0$ reactivity and its performance for in-situ field applications. The discussion is based

117 on the contemporary knowledge of the mechanism of aqueous contaminant removal by Fe$^0$

118 [52-54]. Much of the impetus for this work has come from the work of Noubactep and Caré

119 [34], who have challenged the concept that nano-Fe$^0$ is a strong reducing agent for

120 contaminant reductive transformation.

121 **2    Nanoscale metallic iron or environmental remediation**

122 To date, nano-Fe$^0$ particles have been reported as largely successful for water and soil

123 treatment [11,31,32,55,56]. A wide variety of redox-amenable organic and inorganic species

124 and non-reducible species (e.g. Cd, Zn) have been efficiently treated. Similar to μm and mm-

125 Fe$^0$, adsorption is considered important only for non-reducible species [52-54, 57-60]. For

126 example, Boparai et al. [59] reported that heavy metals are either reduced (e.g. Cu$^{2+}$, Ag$^{2+}$) at,

127 or directly adsorbed (e.g. Zn$^{2+}$, Cd$^{2+}$) onto the Fe$^0$ surface. They further argued that "the

128 controlling mechanism is a function of the standard redox potential of the contaminant".

129 Recent work has however challenged this concept [36,54], which is explained below.

130 **2.1    Contaminant reduction by nano-Fe$^0$**

131 The chemical reaction between $Fe^0$ and redox-amenable aqueous species is considered to

132 involve three steps: (i) direct electron transfer from $Fe^0$ at the metal surface or through a

133 conductive oxide film on $Fe^0$ (direct reduction), (ii) catalyzed hydrogenolysis by the $H/H_2$

134 (indirect reduction mechanism 1), and (iii) reduction by $Fe^{II}$ species resulting from $Fe^0$

135 corrosion (indirect reduction mechanism 2). In this constellation, $H_2$ is supposed to result

136 from $H_2O$ reduction during anoxic iron corrosion [22,61]. However, evidence exists in the

137 literature, e.g. Stratmann and Müller [62], that even under external oxic conditions, $Fe^0$ is

138 oxidized by $H_2O$ (or more precisely by $H^+$) and $O_2$ by $Fe^{II}$ (Table 2). Despite the significant

139 reaction rate exhibited by nano-$Fe^0$ due to its high surface area, such processes are considered

140 to occur (discounting any quantum size effects) independent of particle size.

141 Table 2 summarizes the half reactions for the aqueous oxidation of $Fe^0$ under both anoxic and

142 oxic conditions. Thermodynamically, the major cathodic reaction depends on the availability

143 of molecular $O_2$ ($E^0 = 0.81$ V). In the absence of $O_2$, $Fe^0$ is oxidized by $H^+$ ($E^0 = 0.00$ V). It

144 can therefore be stated that the rate of $Fe^0$ oxidation is dictated by the concentration of

145 dissolved $O_2$, $H^+$ and $H_2O$ in proximity to $Fe^0$ surfaces. Le Chatelier's principle also states that

146 the consumption of $Fe^{II}$ (via oxidation to $Fe^{III}$) will also result in an increase in $Fe^0$ oxidation.

147 The electrode potential of the redox couple $Fe^{II}/Fe^0$ is -0.44 V, a value which is independent

148 of the particle size (nm, μm or mm). The value -0.44 V is considered largely unchanged due

149 to the presence of alloying materials (e.g. low alloy steel, bimetallic systems).

150 As a consequence, statements including "*nano-$Fe^0$ are more reactive than μm-$Fe^0$ and mm-*

151 *$Fe^0$*" are misleading; as the reactivity of $Fe^0$ (discounting quantum size effects), is independent

152 of the particle size. Any enhanced reactivity reported is likely to be due to the significantly

153 high surface area of nano-$Fe^0$ compared to other forms. A second statement "*bimetallic nano-*

154 *$Fe^0$ is more reactive that monometallic nano-$Fe^0$*" is also a qualitative statement, as the

155 reactivity of the materials depends on numerous factors associated with the materials

156 synthesis route and varies depending on the chemistry of the chosen alloying metal. Ideally,

157  comparisons should be made versus standard reference materials using established standard

158  experimental protocols [63], which once established, will significantly improve the design of

159  future field applications.

## 2.2    Limitations of the nano-$Fe^0$ technology

161  The efficiency of nano-$Fe^0$ for aqueous contaminant reduction faces some key issues for in-

162  situ applications in porous media. These challenges include: (i) the strong tendency of

163  aggregation/agglomeration, (ii) the rapid settlement on subsurface solid phases, (iii) the

164  porosity and permeability loss of porous media [23,35,64,65]. Aggregation and settlement

165  limit nano-$Fe^0$ transport through porous media. Porosity and permeability loss limit nano-$Fe^0$

166  transport to target contaminants. It was demonstrated that nano-$Fe^0$ may travel only a few

167  centimetres in porous media from the injection position under typical groundwater conditions

168  [11,12,30,66]. Accordingly, recent efforts have been made to (i) increase the porosity of

169  porous media, (ii) mechanically increase the distribution of nano-$Fe^0$, and/or (iii) chemically

170  modify nano-$Fe^0$ for improved aqueous mobility in porous networks.

## 2.3    Improving the efficiency of the nano-$Fe^0$ systems

### 2.3.1    Dispersion agents

173  Methods to improve the aqueous mobility of nano-$Fe^0$ have received the greatest research

174  interest. It has been determined that the key to improving particle mobility is found in

175  modifying their surface properties such that the nano-$Fe^0$ have significantly improved

176  colloidal stability and a commensurate reduction in the likelihood of adherence to mineral

177  surfaces. Several synthetic methods are now available to produce more mobile nano-$Fe^0$.

178  Efficiently tested dispersants include anionic surface chargers (e.g. polyacrylic acid), non-

179  ionic surfactants, starch, and oil [23,67-70].

### 2.3.2    Bimetallic combinations

181  In recent years, noble metals have been used to increase the reactivity of monometallic nano-

182  $Fe^0$ [21,23,71-73]. As mentioned above, this appears counterintuitive as nano-$Fe^0$ is already high reactivity due to its size [15,26,29,42] and is unstable during synthesis, storage and application [69]. This chemical instability has been documented as a key reason for the observed lower efficiency exhibited by nano-$Fe^0$ systems compared to μm and mm-$Fe^0$ [11]. Accordingly, it is questionable whether further enhancing the reactivity of nano-$Fe^0$, e.g. by plating with more noble elements, may be of any benefit. The reactivity of nano-$Fe^0$ will be discussed in the next section on the basis of mathematical modelling.

## 3    Significance of increased reactivity

### 3.1    The problem

The increased $Fe^0$ reactivity from mm to nm size should be better characterized. The relative reactivity of four different materials is discussed on the basis of 1 kg $Fe^0$: one nm-$Fe^0$ ($d_0 = 25$ nm), one μm-$Fe^0$ ($d_0 = 25$ μm), and two mm-$Fe^0$ ($d_0 = 250$ and $1000$ μm). Calculations for the number particles (N) in 1 kg of each material and the number of layers (N') in each particle are made after the Eq. (1) and Eq. (2) presented in details elsewhere [9,74].

$$N = \frac{M}{\rho_{Fe}.4/3\pi.R_0^3} \tag{1}$$

$$N' = 2.[4/3(\pi R_0^3)]/a^3 \tag{2}$$

where M is the mass of $Fe^0$ (here 1 kg), $\rho_{Fe}$ is the specific weight of Fe ($7,800$ kg/m³), $R_0$ is the initial radius of the Fe particle ($d = 2*R_0$) and a the lattice parameter ($a = 2.866$ Å).

The results are summarized in Table 3. It can be seen that the number of layers of $Fe^0$ in individual particles varies from 87.2 for nano-$Fe^0$ to more than $3*10^6$ for mm-$Fe^0$ ($d = 1$ mm). In the meantime, the number of particles in 1 kg decreased from $1.96*10^{18}$ for nano-$Fe^0$ to only $3.1*10^4$ for mm-$Fe^0$. The ratio of the number of $Fe^0$ layers in each particle to the number of $Fe^0$ layers in nano-$Fe^0$ varies from 1 to $4*10^4$. This ratio corresponds to the relative time (τ) as defined later (section 3.2). On the other hand, the ratio of the number of particles in 1 kg of nano-$Fe^0$ to the number of particles in the same mass of each other materials varies from 1 to

208    $6.4*10^{13}$. These results are summarized in Fig. 1. Instead of the mass of $Fe^0$, the number of

209    electrons released by the conversion of $Fe^0$ to $Fe^{II}$ is used to assess the kinetics of $Fe^0$

210    consumption. This is discussed in the next section.

211    **3.2    Relative corrosion kinetics of $Fe^0$ materials**

212    For the discussion in this section, uniform corrosion for spherical particles is assumed.

213    Individual particles corrode independently until material depletion. It is further assumed for

214    simplicity that individual layers corrode with the same kinetics independent of particle size (d

215    = $2*R_0$). The latter assumption is conservative as larger particles react slower than smaller

216    [24,29,61]. With these assumptions, a relative time ($\tau$) can be defined while taking the time

217    for the corrosion of the smallest particle (here 87.2 layers of nano-$Fe^0$) or $t_{\infty,nano}$ as unit.

218                                $\tau = t/t_{\infty,nano}$                                (3)

219    Accordingly, one unit of time corresponds to the time to nano-$Fe^0$ depletion. Remember that

220    all $1.96*10^{18}$ particles in the 1 kg of nano-$Fe^0$ simultaneously corrode with the same kinetics.

221    The results of the calculations are presented in Fig. 2. From Fig. 2a and Tab. 3 it can be seen

222    that after nano-$Fe^0$ depletion, the material with 1000 µm (or 1 mm) diameter will still react for

223    more than $3*10^4$ times longer than the time necessary for nano-$Fe^0$ depletion ($\tau = 4*10^4$, see

224    Tab. 3). Fig. 2b shows that the mm-$Fe^0$ with 250 µm diameter is depleted after about

225    $10^4*t_{\infty,nano}$.

226    Based on the assumptions above, the service life of a nano-$Fe^0$ particle can be estimated.

227    Table 4 summarizes the results of such estimations while varying the service life of a 1 mm

228    $Fe^0$ particle from 5 to 40 years. This assumption is based on the fact that conventional $Fe^0$

229    walls are supposed to function for several decades (here up to 4 decades). Results show (Tab.

230    4) that the maximum life-span of a nano-$Fe^0$ is about 8.8 hours (less than one day). In other

231    words, following approximately 9 hours from subsurface deployment it is suggested that all

232    nano-$Fe^0$ would be reactively exhausted. The success of this is dependent on three key

233 factors: (i) the hydrodynamic conditions: pore size, porosity, flow velocity and degree of

234 mixing or turbulence, (ii) the water chemistry and the affinity of nano-$Fe^0$ and its

235 transformation products to the soil materials, and (iii) the reactivity of $Fe^0$.

236 It is certain, that the dynamic process of transformation of concentric layers of $Fe^0$ atoms to

237 concentric layers of iron (hydr)oxides can not be linear [34]. In fact, effects similar to "case

238 hardening" for food- and wood-drying will lead to "surface hardened layers" [75, 76] leading

239 to differential kinetics/extents of $Fe^0$ passivation for different particle size ranges. In other

240 words, the extend of restricted corrosion rates through resulting surface hardened layers will

241 be different for nm-, μm- and mm-$Fe^0$. Bearing this in mine, the very short relative life-span

242 of a nano-$Fe^0$ estimated above will be used for the discussion in this work. It is certain that

243 "case hardening"-like effects will prolong this hypothetical life-span to some days or weeks.

244 **3.3    Extent of iron corrosion from $Fe^0$ materials**

245 A discussion as to the extent of $Fe^0$ consumption is limited in the present section to $\tau = 1$ or

246 $t_{\infty,nano}$. It is considered for simplification that the sole iron corrosion product is $Fe_3O_4$. The

247 corresponding coefficient of volumetric expansion is $\eta_{Fe3O4} = 2.08$ (Eq. 4) [46]. Using $\rho =$

248 M/V, the volume of Fe corresponding to 1 kg $Fe^0$ is calculated as 127.0 mL ($V_0$). This is the

249 initial volume of $Fe^0$ ($V_0$). Following corrosion, this volume is partly or totally consumed.

250 The volume ($\Delta V$) corresponding to the volume of pores occupied by the volumetric expansion

251 of corrosion products can be estimated.

252 Assuming that the coefficient of volumetric expansion  ($\eta$) ("rust expansion coefficient" or

253 "specific volume") [45-47] of the reaction products is:

254 $$\eta = V_{oxide}/V_{Fe} \tag{4}$$

255 where $V_{oxide}$ is the volume of the reaction product and $V_{Fe}$ the volume of the parent $Fe^0$.

256 The volume $\Delta V$ characterizing the extent of porosity loss due to volumetric expansion is

257 given by Eq. 5:

258          $$\Delta V = (\eta - 1) * V_{consumed\ Fe} \qquad (5)$$

259          $$\Delta V = V_\infty - V_0 = \nu*(\eta_{Fe3O4} - 1)*V_0 \qquad (5a)$$

260   Where $V_{consumed\ Fe}$ is the volume of consumed $Fe^0$ at time $t_\infty$, $V_0$ is the volume occupied by the

261   initial $Fe^0$ particles and $\nu$ ($\nu \leq 1$) is the fraction of the initial amount of $Fe^0$ (1 kg) which has

262   reacted at $t_{\infty,nano}$. $V_\infty$ is the total volume occupied by residual $Fe^0$ and in-situ formed corrosion

263   products. $t_\infty$ corresponds to nano-$Fe^0$ depletion (25 nm in this section). In the discussion on

264   the reactivity at nano-scale, $t_\infty$ corresponds to the depletion of the material with 10 nm

265   diameter (section 4).

266          $$V_\infty = \eta * V_{consumed\ Fe} + (V_0 - V_{consumed\ Fe}) \qquad (6)$$

267          $$V_\infty = V_0\ [1 + \nu*(\eta_{Fe3O4} - 1)] \qquad (6a)$$

268   The volumetric expansion ($\Delta V$, Eq. 5) can be characterized as percent of the initial volume

269   ($V_0$) using Eq. (7):

270          $$\Delta V\ (\%) = 100*\nu*(\eta_{Fe3O4} - 1) \qquad (7)$$

271   Table 5 summarizes the results. It is shown that at $\tau = 1$ (nano-$Fe^0$ depletion), a volume

272   augmentation of 108 % has occurred in the nano-$Fe^0$ system, with volume augmentations in

273   all other systems lower than 0.5 %. This clearly shows that the porosity of the subsurface will

274   be significantly influenced by nano-$Fe^0$ at $t_\infty$. Remember that 100 % reactive exhaustion of

275   nano-$Fe^0$ is predicted to occur by approximately 9 hours time. During this same period the

276   porosity loss due to expansive iron corrosion is likely to be negligible for all other $Fe^0$ particle

277   size fractions. Calculations for Akageneite $\beta$-FeOOH ($\eta_{FeOOH} = 3.48$) as sole corrosion

278   products shows that $V_{\infty,nano} = 448.7$ mL, $\Delta V = 320.5$ mL or 250.4 %. The examples of $Fe_3O_4$

279   (anoxic) and FeOOH (oxic) demonstrate the crucial importance of the nature of formed

280   corrosion products for the discussion of the extent of porosity loss.

281   Another important aspect of $Fe^0$ consumption is given by the number of moles of $Fe^0$ that

282   have been oxidized (Tab. 5). Assuming contaminant reduction, Tab. 5 shows that after $\tau = 1$,

283    35.71 moles of electrons have been released in the nano-$Fe^0$ system but less than 0.11 moles

284    in all other systems. In other words, up to 35.71 moles of electrons are available for

285    contaminant reduction per kg nano-$Fe^0$ within a few hours of reaction (< 9 hours). But what

286    proportion of the electrons produced would reach the contaminant within this period? That is

287    the major question to be answered for the further development of the nano-$Fe^0$ technology for

288    in-situ applications.

289    **4      Reactivity of nano-$Fe^0$ materials**

290    The presentation until now has discussed the reactivity of nano-$Fe^0$ in comparison to larger

291    scale $Fe^0$. Section 4 will focus only on the nanoscale size fraction (d ≤ 100 nm). Equations 1-7

292    will be used and the particle size will vary from 10 to 100 nm. As stated above $t_\infty$ is for a

293    nano-$Fe^0$ of 10 nm diameter and the reaction proceeding until 100 % reactive exhaustion has

294    been achieved.

295    **4.1    $Fe^0$ reactivity at nanoscale**

296    Table 6 summarizes the results of calculations for the number of $Fe^0$ particles and number of

297    layers of $Fe^0$ in each nano-$Fe^0$. It is shown that 1 kg of the material with d = 100 nm contains

298    1000 times more particles than a material of d = 10 nm.

299    Table 6 also shows that the maximum value of the relative time (τ) is 10 (or $10^1$). This is more

300    practical for graphical representations than situations where nano-$Fe^0$ are compared to larger

301    particles (t ≤ $10^4$). The physical significance of τ is more important, it means that if a nano-

302    $Fe^0$ with a diameter of 10 nm depletes after 2 days, the material with a diameter 100 nm will

303    deplete after 20 days. For field applications the selection of the particle size to be used should

304    be dictated by site specific characteristics. Which diameter could quantitatively reach the

305    contaminants before depletion? And what fraction of the material will have already oxidized

306    on the path? What is the impact of this oxidation on the transport of nano-$Fe^0$ in the porous

307    aquifer? These are some key questions to be answered in order to give this possibly very

308    efficient technology a scientific basis.

309    Fig. 3 summarizes the evolution of the volumetric expansion in all 5 nano-Fe$^0$ systems. It can

310    be seen from Fig. 3a that the smallest material (d = 10 nm) experiences the 108 % volumetric

311    expansion within a short time ($\tau$ = 1) while the larger materials (d = 100 nm) needs 10 more

312    time for the same expansion. Accordingly, beside the question whether the material will reach

313    the contaminant under site specific conditions, the question has to be answered how the

314    volumetric expansion will impact the aquifer porosity (and permeability).

315    Fig. 3b compares the variation of the volumetric expansion for two different iron corrosion

316    products, $Fe_3O_4$ and $FeOOH$, which are considered the most likely products in anoxic and

317    oxic aquifers respectively. It also shows that when designing a nano-Fe$^0$ injection strategy,

318    however, the availability of oxidizing species (e.g. $MnO_2$, $O_2$) must also to be taken into

319    account. Fig. 3a shows that under both conditions the trend of porosity loss is similar but the

320    extent is proportional to the coefficient of volumetric expansion ($\eta$). In particular, at $\tau$ = 1, the

321    system with the material d = 10 nm experiences 250 % volumetric expansion under oxic

322    conditions and only 110 % under anoxic conditions. As a result the kinetics of more rapid Fe$^0$

323    corrosion in an oxygen-rich environment must also be considered for an effective treatment

324    strategy.

325    **4.2    Fe$^0$ reactivity of nano-bimetallics**

326    The reactivity of monometallic nano-Fe$^0$ can be improved by combining it with a noble metal.

327    Assuming $\alpha$ ($\alpha$ > 1) the coefficient of reactivity enhancement, the relation between the

328    relative time of a bimetallic system ($\tau_{Fe/M}$) and that of a non plated metal ($\tau_{Fe}$) is given by Eq.

329    6:

330                    $\tau_{Fe} = \alpha * \tau_{Fe/M}$                (8)

331    To characterize the impact of plating on nano-Fe$^0$, the material with the largest size (d = 100

332    nm) will be plated by three hypothetical metals ($M^0_1$, $M^0_2$ and $M^0_3$) to yield a reactivity factor

333    of 2.5 (for Fe$^0$/$M^0_1$), 5 (for Fe$^0$/$M^0_2$) and 10 (for Fe$^0$/$M^0_3$). The considered $\alpha$ values of ($\alpha \leq 10$)

334  are realistic and even conservative. In fact, reported reactivity enhancement is essentially

335  larger [22,77]. For example, Zhuang et al. [77] reported that palladized nano-$Fe^0$ promoted

336  the dehalogenation kinetics for polybrominated diphenyl ethers by orders of magnitude equal

337  to 2, 3 and 4 ($\alpha \geq 100$). The results of the calculations for the four systems (d = 10 nm) are

338  summarized in Fig. 4. The system with d = 10 nm is represented for comparison. It can be

339  seen that system $Fe^0/M^0_3$ (d = 100 nm) is as reactive as $Fe^0$ (d = 10 nm). Given that the

340  reactivity of nano-$Fe^0$ (d = 100 nm) could already significantly been too high in some

341  situations, the results from Fig. 4 strongly question the suitability of plating at nano-scale.

342  Accordingly, while the application of bimetallic $Fe^0$ is definitively useful at μm- and mm-

343  scale, it usefulness at nano-scale is likely inappropriate. It can also be noted that by increasing

344  the reactivity of the material the rate at which volumetric pore clogging also increases. As a

345  consequence it should be acknowledged that there exists a conceptual play-off between

346  increased reaction rate and increased porosity loss, the impact of which will vary depending

347  on the physiochemical conditions of each contaminated site.

348  **4.3    Characterizing the process of reactivity loss**

349  To better characterize the process of porosity loss due to the volumetric expansion of nano-

350  $Fe^0$, the evolution of the porosity of a sand column filled with nano-$Fe^0$ will be discussed as

351  volumetric expansion proceeds. A laboratory column with a height h (h = 75.0 cm) and

352  diameter D (D = 5.0 cm) is composed of spherical sand particles (d = 5.0 mm). The

353  compactness of the column is ideally C = 0.64 [9,74]. The pore volume is given by Eq. 9:

354  $$V_{pore} = V * (1 - C) \tag{9}$$

355  where V is the apparent volume of the sand column (V = h * π *$D^2$/4).

356  It is supposed that the nano-$Fe^0$ particles fill the inter-granular porosity of the sand column

357  $V_{pore}$ without modifying the compactness C and the apparent volume V of the sand column.

358  The residual porosity of the sand column ($V'_{pore}$) is given by Eq. 10:

359  $$V'_{pore} = V*(1-C) - V_0 \tag{10}$$

360    where $V_0$ is the volume of the Fe particles.

361    The evolution of the residual porosity ($V'_{pore}$) as nano-$Fe^0$ particles undergo volumetric

362    expansive corrosion is considered by introducing the specific volume ($\eta$) of the reaction

363    products according to Eq. 11:

364    $$V'_{pore}= V*(1 - C) - (V_0 - V_{consumed\ Fe}) - \eta * V_{consumed\ Fe} \qquad (11)$$

365    $$V'_{pore}= V*(1 - C) - V_0 - (\eta-1) * V_{consumed\ Fe} \qquad (11a)$$

366    where $V_{consumed\ Fe}$ is the volume of nano-$Fe^0$ particles which is consumed at a given time.

367    Equations 9 through 11 are very useful to design reactive zone. However, they are limited to

368    describe the initial ($V_{pore}$) and a final conditions ($V'_{pore}$) regardless on the nature of iron

369    corrosion products and the kinetics of the process.

370    Using a sand column comparable to one of those used by Moraci and Calabrò [78] and 1 kg of

371    nano-$Fe^0$, the process of pore filling (porosity loss) can be better characterized. For simplicity

372    nano-$Fe^0$ considered as transported by a biodegradable dispersant which does not significantly

373    contribute to porosity loss. As shown in section 3.3, 1 kg of nano-$Fe^0$ occupies a volume of

374    127 mL. The initial pore volume of the sand column calculated after Eq. 9 is 530.36 mL (100

375    % porosity), i.e. a capacity for approximately 4.17 kg of nano-$Fe^0$. Filling the initial pore

376    volume of the sand column (530.36 mL) with 1 kg of nano-$Fe^0$ (127.00 mL) yields a 23.9 %

377    porosity loss (Tab. 7). This, however, does not take into account the expansive nature of iron

378    during oxidative corrosion.

379    Using the 8 possible iron corrosion products documented by Caré et al. [46] and their

380    respective coefficient of volumetric expansion ($2.08 \leq \eta \leq 6.40$), the extent of porosity loss is

381    calculated and summarized in Table 7. The results show that the residual volume of pores

382    ($V'_{pore}$) decreases with increasing $\eta$ values and is zero for $Fe(OH)_3$ and $Fe(OH)_3.3H_2O$ (100

383    % porosity loss). Ferrihydrite ($Fe(OH)_3.3H_2O$) is the largest known iron corrosion products.

384    In other words, depending on environmental conditions as little as 1 kg of nano-$Fe^0$ could

385    clog the tested column. Although this discussion considers the nature of the corrosion

386 products, there are other important factors which must be considered. The negative values (-

387 3.04 and -282.4 mL) corresponds to the mass of $Fe^0$ which will not oxidized because of lack

388 of space for expansion [9,74].

389 The extent of porosity loss ($\Delta V$ in %) given in Tab. 7 assumes uniform distribution of nano-

390 $Fe^0$ in the whole column. This is, however, not a very good field representation. For example,

391 if 1 kg of nano-$Fe^0$ ($V_0 = 127$ mL) is uniformly distributed only in the first third of the column

392 ($V^{1/3}_{pore} = 176.8$ mL), with $Fe_3O_4$ as the primary corrosion product ($\Delta V = 137.16$ mL) a 78 %

393 porosity loss can be expected. For all other oxide phases it is calculated that complete porosity

394 loss (100 %) will precede nano-$Fe^0$ reactive exhaustion. However, in the practice a system

395 with 78 % porosity loss is considered as clogged. One possibility to avoid the clogging of the

396 entrance zone of a porous system is to intermittently inject calculated amounts of nano-$Fe^0$.

397 The volume to be injected at each event and the time scale between two injections are

398 necessarily determined by site specific characteristics (e.g. aquifer porosity, water flow rate).

399 **5    Discussion**

400 A primary reason behind the interest into the use of nano-$Fe^0$ particles over $\mu$m-$Fe^0$ and mm-

401 $Fe^0$ particles for water treatment is ascribed to a significant increase the materials efficiency

402 [11,12]. For example, as reported by Vodyanitskii [79], Kanel et al. [80] reported near-total

403 remediation of a 1 mg $L^{-1}$ $As^V$ solution within only 10 min by a nano-$Fe^0$ with a specific

404 surface of 24 $m^2/g$, whereas the same goal was achieved by mm-$Fe^0$ (1–2 $m^2/g$) after only 4

405 days or 5760 min (ratio of time 570; average ratio of surface 16). However, this experimental

406 evidence is highly qualitative as neither the number of atoms directly accessible at the surface

407 nor the intrinsic reactivity of individual materials are considered in both cases [34]. For a

408 better comparative result, the following three key conditions must be considered: (i) the

409 intrinsic $Fe^0$ reactivity should be characterized, (ii) the amount of used materials should be

410 based on the reaction stoichiometry, and (iii) the experimental conditions should be relevant

411 for field applications. In particular, the driving force for the transport of contaminants and Fe

412 species should be relevant for field situations: (i) mixing operation (type and intensity) in
413 batch studies, (ii) flow rate and column dimensions in column studies [63].

414 **5.1    Transport of nano-$Fe^0$ to the contaminants**

415 The efficiency of nano-$Fe^0$ for the in-situ treatment of a contaminated aquifer body is
416 intrinsically linked to the extent of physical contact between $Fe^0$ and any aqueous
417 contaminant species present. In some circumstances, contaminants could diffuse to the
418 suspended $Fe^0$ particles and be degraded in the aqueous phase. However, typically the
419 suspended $Fe^0$ particles must migrate to the contaminants. As $Fe^0$ particles are transported
420 from the injection zone to the target contaminant plume by natural groundwater, diffusion
421 experiments under relevant groundwater velocity, using site specific aquifer materials are
422 essential in order to effectively assess the suitability of nano-$Fe^0$ for in-situ applications [81].

423 Contaminants are typically partitioned between sediment and water phases in a "pseudo-
424 equilibrium" state. Therefore, it is likely that $Fe^0$ particles whilst acting to reduce any soluble
425 contaminants are also likely to promote the dissolution of a range of adsorbed chemical
426 species (Le Chatelier's principle). However, as water is also a redox-amenable species the
427 specific mechanism for nano-$Fe^0$ reactivity in a range of conditions is difficult to resolve [30].
428 In other words, nano-$Fe^0$ is readily oxidized by $H_2O$ during subsurface migration to the target
429 contaminant plume and also competes with any other redox-amenable (including
430 contaminants) present in the groundwater. Additionally, expansive iron corrosion will yield
431 voluminous iron (hydr)oxides (Tab. 7) with limited aqueous mobility due to (i) an increased
432 size and weight, and (ii) a possible increased affinity to aquifer material.

433 This discussion has intentionally neglected the segregation between parent compounds, the
434 reaction products and their relative affinity to $Fe^0$ and Fe (hydro)oxides. The fact that the core
435 $Fe^0$ is always covered by oxide layers is also neglected for simplification. The process of
436 preferential flow which is crucial in predicting mass transfer in the subsurface is also not
437 considered [82-84]. However, it is clearly shown, that due the acute redox sensitivity of nano-

438    $Fe^0$ and the subsequent significant formation of highly voluminous oxidative corrosion

439    products it is likely that for environmentally relevant distances (m), a significant proportion of

440    the originally injected nano-$Fe^0$ will remain "clogged" in pore spaces.

441    Beside the transport of nano-$Fe^0$ to contaminants, the possibility of quantitative contaminant

442    desorption and their subsequent transformation by suspended $Fe^0$ could be considered.

443    However, it is not likely that concentration-gradient-driven mass transfer could be

444    quantitative at considered distances (m). It should be recalled that the slow kinetics of

445    contaminant desorption form aquifer materials is the major cause of the ineffectiveness of the

446    pump-and-treat technology for groundwater remediation [87-89].

447    This section has shown that it is likely that the success of nano-$Fe^0$ for in-situ remediation is

448    seriously limited by the intrinsic formation of voluminous iron corrosion products

449    [11,12,30,39,40]. Bearing this in mind, the next section suggests an alternative nano-$Fe^0$

450    subsurface deployment mechanism that more effectively takes into account the

451    aforementioned nano-$Fe^0$ hydraulic mobility issues than conventional injection processes: the

452    formation of a nano-$Fe^0$ "redox-front" injection array system for progressive contaminant

453    reduction. The geochemical process of redox-front migration is a well-documented one

454    [83,90,91].

455    ## 5.2    Nano-$Fe^0$ as source of $Fe^{II}$ for a redox-front?

456    ### 5.2.1  The concept

457    The progressive consumption of mm-$Fe^0$ (Fig. 1; Tab. 3) is the guarantee for the long-term

458    efficiency of reactive barriers [11]. In fact, continuously generated small amount of high

459    reactive iron minerals [36,52-54,92-95] are sufficient for the removal of contaminants which

460    are present in trace amounts [96]. As discussed above, for nano-$Fe^0$ however, (i) $Fe^0$ reactive

461    exhaustion typically occurs in a relatively short time scale (< 9 hours) and, (ii) it is likely that

462    nano-$Fe^0$ subsurface mobility is significantly retarded or even prevented due to the volumetric

463    expansive nature of iron corrosion [46]. As a consequence an alternative method of

464     subsurface deployment is suggested in the current work: the deployment of a linear nano-$Fe^0$

465     injection array oriented perpendicular to the flow direction of the contaminant plume. The

466     injected nano-$Fe^0$ can effectively form a redox-front (roll-front) which migrates through the

467     contaminated zone and transforms the contaminants during its migration as illustrated in Fig.

468     5. The $Fe^{II}/Fe^{III}$ roll-front travels across the contaminated zone with all possible mechanisms

469     (e.g. diffusion, dispersion, convection, preferential flow) and the contaminants are

470     transformed and immobilized during the cycle $Fe^{II} \Leftrightarrow Fe^{III}$. In other words, it is a plume of

471     $Fe^{II}/Fe^{III}$ formed from injected nano-$Fe^0$ which migrates through the contaminated zone and

472     "sweeps" the contaminants. As a consequence this method considers all nano-$Fe^0$ mobility

473     issues.

### 474    5.2.2    Nano-$Fe^0$ as $Fe^{II}$ generator

475     Nano-$Fe^0$ in the aqueous phase is certainly a $Fe^{II}/Fe^{III}$ producer. $Fe^{II}$-species are the main

476     reducing agents for contaminants under both anoxic and oxic conditions [62]. The reducing

477     capacity of $Fe^{II}$-species nominally depends on the pH value [54,77,97,98]. Microbial activity

478     could regenerate $Fe^{II}$ (bio-corrosion) for more contaminant reduction [79]. In this case, more

479     contaminant is reduced than can be predicted from the reaction stoichiometry. In order words,

480     the operating mode of nano-$Fe^0$ for contaminant reduction can be summarized as follows: (i)

481     $Fe^0$ is oxidized to produce $Fe^{II}$, (ii) $Fe^{II}$ reduces the contaminant and is oxidized to $Fe^{III}$, and

482     (iii) a proportion of $Fe^{II}$ is regenerated by the biological reduction of $Fe^{III}$. Accordingly,

483     before $Fe^0$ depletion, there are three sources of $Fe^{II}$: (i) the $Fe^0$ mediated abiotic oxidation by

484     $H_2O$, (ii) the $Fe^0$ mediated abiotic oxidation by $Fe^{III}$, and (iii) the biological reduction of $Fe^{III}$.

485     After $Fe^0$ depletion, the only remaining source of $Fe^{II}$ is the biological reduction of $Fe^{III}$.

486     Provided that the appropriate micro-organism species are present in the subsurface, this

487     process, however, could conceptually proceed for a significantly long time period [35].

488     Evidence suggests that such micro-organism colonies can be sustained by a consistent supply

489     of $Fe^{II}$, $Fe^{III}$ and molecular hydrogen ($H/H_2$). Another further process that is worth noting is

490    the generation of atomic or molecular hydrogen ($H/H_2$) by $Fe^0$ mediated hydrolysis reactions,

491    which is likely to aid and the aforementioned biotic processes [35].

492    The abiotic conversion of $Fe^{III}$ to $Fe^{II}$ has been successfully utilised in the hydrometallurgy

493    industry, for example Lottering et al. [99] reported on the sustainable use of $MnO_2$ for the

494    abiotic regeneration of $Fe^{III}$ for $U^{IV}$ oxidation.

495    The fate of contaminant reduction products is discussed in the next section.

### 5.3    Mechanism of contaminant removal by injected nano-$Fe^0$

497    The successful application of nano-$Fe^0$ injection technology for in-situ remediation is highly

498    dependent on a comprehensive understanding of the fundamental processes governing the

499    processes of contaminant removal. The hitherto discussion has focused on reductive

500    transformations by nano-$Fe^0$. However, contaminant reductive transformation is not a

501    guarantee for contaminant removal [52-54]. Additionally, certain reaction products are more

502    toxic than their parent compounds [100]. Accordingly, efforts have to be focused on the

503    specific mechanism of aqueous contaminant removal. Relevant removal processes include: (i)

504    adsorption, (ii) chemical precipitation, (iii) co-precipitation, (iv) size exclusion or straining,

505    and (v) volatilization [52-54,79,101-104]. Chemical precipitation is a characteristic of

506    inorganic compounds when the solubility limit is exceeded [101,102,105]. Volatilization is

507    subsequent to chemical transformation yielding gaseous species like $AsH_3$, $CH_4$, $CO_2$, $H_2$, $N_2$.

508    In $Fe^0$ reactive barrier systems, contaminants are efficiently removed by the combination of

509    adsorption, co-precipitation and size exclusion within the engineered barrier [14,106]. As a

510    result the current discussion concentrates on such processes..

511    With $Fe^0$ ($< 1$ $m^2/g$) first transformed to voluminous hydroxides species ($> 100$ $m^2/g$) and

512    subsequently transformed to oxides ($< 40$ $m^2/g$), contaminant size exclusion (straining) is

513    driven by the dynamic cycle of expansion/compression accompanying the corrosion process

514    [14,107]. During these cycles contaminants are enmeshed and sequestrated in a "matrix" of

515    iron corrosion products.

516  For conventional nano-$Fe^0$ injection arrays, size exclusion may play an important role (i) in

517  proximity for $Fe^0$ particles, and (ii) by reducing the pore space during expansive corrosion of

518  the materials. However, if roll-fronts are formed as discussed above, the extent of

519  permeability loss in aquifer will be limited. The roll-front could act as a colloidal reactive

520  barrier for the removal of parent contaminants and reaction products. Species are removed or

521  immobilized by colloids and not because they are reduced. More research is needed to test

522  this hypothesis.

523  ### 5.3.1   $Fe^{II}/Fe^{III}$ redox-front as a colloidal reactive barrier

524  Aqueous contaminants have been reported to be quantitatively removed both during abiotic

525  and biotic (i) oxidation of $Fe^{II}$ and (ii) reduction of $Fe^{III}$ [107-109]. On the other hand,

526  injection of $Fe^{III}$ salts for adsorptive contaminant removal has been reported [110,111].

527  Accordingly, the migration of the $Fe^{II}/Fe^{III}$-redox-front may be coupled to quantitative

528  contaminant removal by adsorption and co-precipitation.

529  The primary reason for contaminant removal during these redox reactions is the colloidal

530  nature of in-situ generated Fe species [$Fe(OH)_2$, $Fe(OH)_3$] [112], which necessarily

531  experience volumetric contraction to form oxides (of $Fe^{II}$ or $Fe^{III}$). Contaminants are first

532  adsorbed by highly reactive colloids and are co-precipitated during transformation to

533  amorphous and crystalline oxides [79,113]

534  ## 6      Concluding remarks

535  Constructed geochemical barriers of metallic iron ($Fe^0$) have been used for groundwater

536  remediation since 1996 [1,5,6,79,114]. In recent years, however, nano-$Fe^0$ has received

537  proclaim as a new tool for water treatment due to (i) improvements in reactivity and

538  associated aqueous contaminant removal performance compared to conventional materials,

539  and (ii) the option of subsurface deployment via injection for targeted in-situ treatment of

540  contaminant plumes [11,12].

541 Considering reactivity first, the current work has highlighted the need for prudent use of
542 terminology. Discounting any quantum size effects, which are only prevalent for $Fe^0$ less than
543 approximately 10 nm in diameter, the reactivity of nano-$Fe^0$ as a function of surface area is no
544 more reactive than larger forms. Nano-$Fe^0$ only exhibits such high reactivity due to it
545 significantly high surface area as a function of mass/volume. Despite this, a recent trend in
546 research has been the development of bimetallic nano-$Fe^0$ wherein the combination of a noble
547 metal acts to further increase the reactivity of nano-$Fe^0$. It is argued in the current work that as
548 reactive exhaustion is already achieved by monometallic nano-$Fe^0$ in the order of minutes this
549 seems counterintuitive for the majority of environmental applications.

550 Considering the nano-$Fe^0$ subsurface injection procedure, in the current work it has been
551 highlighted that the hydraulic mobility of the particles is likely to be significantly retarded by
552 voluminous expansion due to particle corrosion. An alternative nano-$Fe^0$ injection procedure
553 has been suggested herein. The injected nano-$Fe^0$ effectively forms an in-situ migrating front
554 which possibly reductively transforms contaminant and removes reduced species by
555 adsorption and co-precipitation.

556 It is also outlined in the current work that a number of studies with experiments "proclaimed"
557 as analogous to environmental systems are largely overlooked a range of operational drivers
558 including changes in nano-$Fe^0$ (i) reactivity, and (ii) voluminous as a function of time. It is
559 hoped that the huge literature on redox-front migration [115-118] and the cycle of iron in the
560 hydrosphere ([79] and ref. therein) will now be used for the further development of nano-$Fe^0$
561 injection technology.

562 **Acknowledgments**

563 Mohammad A. Rahman (Angewandte Geologie - Universität Göttingen) is acknowledged for
564 technical support.

**Cited References**

566 [1] R.W. Gillham, S.F O'Hannesin, Enhanced degradation of halogenated aliphatics by zero-
567 valent iron, Ground Water 32 (1994) 958–967.

568 [2] L.J. Matheson, P.G. Tratnyek, Reductive dehalogenation of chlorinated methanes by iron
569 metal, Environ. Sci. Technol. 28 (1994) 2045–2053.

570 [3] D.W. Blowes, C.J. Ptacek, J.L. Jambor, In-situ remediation of Cr(VI)-contaminated
571 groundwater using permeable reactive walls: laboratory studies, Environ. Sci. Technol.
572 31 (1997) 3348–3357.

573 [4] S.F. O'Hannesin, R.W. Gillham, Long-term performance of an in situ "iron wall" for
574 remediation of VOCs, Ground Water 36 (1998) 164–170.

575 [5] M.M. Scherer, S. Richter, R.L. Valentine, P.J.J. Alvarez, Chemistry and microbiology of
576 permeable reactive barriers for in situ groundwater clean up, Rev. Environ. Sci.
577 Technol. 30 (2000) 363–411.

578 [6] A.D. Henderson, A.H. Demond, Long-term performance of zero-valent iron permeable
579 reactive barriers: a critical review, Environ. Eng. Sci. 24 (2007) 401–423.

580 [7] G. Bartzas, K. Komnitsas, Solid phase studies and geochemical modelling of low-cost
581 permeable reactive barriers, J. Hazard. Mater. 183 (2010) 301–308.

582 [8] Li L., Benson C.H., Evaluation of five strategies to limit the impact of fouling in
583 permeable reactive barriers, J. Hazard. Mater. 181 (2010) 170–180.

584 [9] C. Noubactep, S. Caré, Dimensioning metallic iron beds for efficient contaminant
585 removal, Chem. Eng. J. 163 (2010) 454–460.

586 [10] J.Y. Kim, H.-J. Park, C. Lee, K.L. Nelson, D.L. Sedlak, J. Yoon, Inactivation of
587 escherichia coli by nanoparticulate zerovalent iron and ferrous ion, Appl. Environ.
588 Microbiol. 76 (2010) 7668–7670.

589   [11] S. Comba, A. Di Molfetta, R. Sethi, A comparison between field applications of nano-,

590        micro-, and millimetric zero-valent iron for the remediation of contaminated aquifers,

591        Water Air Soil Pollut. 215 (2011) 595–607.

592   [12] M. Gheju, Hexavalent chromium reduction with zero-valent iron (ZVI) in aquatic

593        systems, Water Air Soil Pollut. (2011) doi 10.1007/s11270-011-0812-y.

594   [13] S.-W. Jeen, R.B. Gillham, A. Przepiora, Predictions of long-term performance of

595        granular iron permeable reactive barriers: Field-scale evaluation, J. Contam. Hydrol.

596        123 (2011) 50–64.

597   [14] C. Noubactep, Metallic iron for safe drinking water production, Freiberg Online

598        Geology, 27 (2011) 38 pp, ISSN 1434-7512. (www.geo.tu-freiberg.de/fog)

599   [15] C.-B. Wang, W.-x. Zhang, Synthesizing nanoscale iron particles for rapid and complete

600        dechlorination of TCE and PCBs, Environ. Sci. Technol. 31 (1997) 2154–2156.

601   [16] S.M. Ponder, J.G. Darab, T.E. Mallouk, Remediation of Cr(VI) and Pb(II) aqueous

602        solutions using supported, nanoscale zero-valent iron, Environ. Sci. Technol. 34 (2000)

603        2564–2569.

604   [17] R. Muftikian, Q. Fernando, N. Korte, A method for the rapid dechlorination of low

605        molecular weight chlorinated hydrocarbons in water, Water Res. 29 (1995) 2434–2439.

606   [18] N.E. Korte, J.L. Zutman, R.M. Schlosser, L. Liang, B. Gu, Q. Fernando, Field

607        application of palladized iron for the dechlorination of trichloroethene, Waste Manage.

608        20 (2000) 687–694.

609   [19] B. Schrick, J.L. Blough, A.D. Jones, T.E. Mallouk, Hydrodechlorination of

610        trichloroethylene to hydrocarbons using bimetallic nickel–iron nanoparticles. Chem.

611        Mater. 14 (2002) 5140–5147.

612   [20] B. Karn, T. Kuiken, M. Otto, Nanotechnology and in situ remediation: A review of the

613        benefits and potential risks. Environ. Health Perspectives 117  (2009) 1832–1831.

614     [21] V. Nagpal, A.D. Bokare, R.C. Chikate, C.V. Rode, K.M. Paknikar, Reductive
615        dechlorination of ?-hexachlorocyclohexane using Fe–Pd bimetallic nanoparticles, J.
616        Hazard. Mater. 175 (2010) 680–687.

617     [22] K.-F. Chen, S. Li, W.-x. Zhang, Renewable hydrogen generation by bimetallic zerovalent
618        iron nanoparticles, Chem. Eng. J. (2011), doi:10.1016/j.cej.2010.12.019.

619     [23] S. Mossa Hosseini, B. Ataie-Ashtiani, M. Kholghi, Nitrate reduction by nano-Fe/Cu
620        particles in packed column, Desalination (2011) doi:10.1016/j.desal.2011.03.051.

621     [24] W.-X. Zhang, C.-B. Wang, H.-L. Lien, Treatment of chlorinated organic contaminants
622        with nanoscale bimetallic particles, Catal. Today 40 (1998) 387–395.

623     [25] R.W. Gillham, Discussion of Papers/Discussion of nano-scale iron for dehalogenation.
624        by Evan K. Nyer and David B. Vance (2001), Ground Water Monitoring &
625        Remediation, v. 21, no. 2, pages 41–54, Ground Water Monit. Remed 23 (2003) 6–8.

626     [26] W.-x. Zhang, Nanoscale iron particles for environmental remediation: an overview, J.
627        Nanopart. Res. 5 (2003) 323–332.

628     [27] X.Q. Li, D.W. Elliott, W.X. Zhang, Zero-valent iron nanoparticles for abatement of
629        environmental pollutants: materials and engineering aspects, Crit. Rev. Solid State
630        Mater. Sci. 31 (2006) 111–122.

631     [28] C. Macé, Controlling groundwater VOCs: do nanoscale ZVI particles have any
632        advantages over microscale ZVI or BNP? Pollut. Eng. 38 (2006) 24–27.

633     [29] C. Macé, S. Desrocher, F. Gheorghiu, A. Kane, M. Pupeza, M. Cernik, P. Kvapil, R.
634        Venkatakrishnan, W.-X. Zhang, Nanotechnology and groundwater remediation: A step
635        forward in technology understanding, Remed. J. 16 (2006) 23–33.

636     [30] P.G. Tratnyek, R.L. Johnson, Nanotechnologies for environmental cleanup, Nano Today
637        1 (2006) 44–48.

638     [31] T. Pradeep, Anshup, Noble metal nanoparticles for water purification: A critical review,
639        Thin Solid Films 517 (2009) 6441–6478.

640 [32] A. Agarwal, H. Joshi, Environmental sciences application of nanotechnology in the
641 remediation of contaminated groundwater: A short review, Recent Res. Sci. Technol. 2
642 (2010) 51–57.

643 [33] N. Müller, B. Nowack, Nano zero valent iron – THE solution for water and soil
644 remediation?. Report of workshop held in Zurich (Switzerland), November 24th 2009
645 (2010). http://www.observatorynano.eu/project/filesystem/files/nZVI_final_vsObservatory.pdf. (Access 2011/04/24)

646 [34] C. Noubactep, S. Caré, On nanoscale metallic iron for groundwater remediation, J.
647 Hazard. Mater. 182 (2010) 923–927.

648 [35] L.G. Cullen, E.L. Tilston, G.R. Mitchell, C.D. Collins, L.J. Shaw, Assessing the impact
649 of nano- and micro-scale zerovalent iron particles on soil microbial activities: Particle
650 reactivity interferes with assay conditions and interpretation of genuine microbial
651 effects, Chemosphere 82 (2011) 1675–1682.

652 [36] C. Noubactep, Comment on "Reductive dechlorination of g-hexachloro-cyclohexane
653 using Fe–Pd bimetallic nanoparticles" by Nagpal et al. [J. Hazard. Mater. 175 (2010)
654 680–687], J. Hazard. Mater. (2011) doi:10.1016/j.jhazmat.2011.03.081.

655 [37] J.R. Peralta-Videa, L. Zhao, M.L. Lopez-Moreno, G. de la Rosa, J. Hong, J.L. Gardea-
656 Torresdey, Nanomaterials and the environment: A review for the biennium 2008–2010,
657 J. Hazard. Mater. 186 (2011) 1–15.

658 [38] Z. Shi, J.T. Nurmi, P.G. Tratnyek, Effects of nano zero-valent iron on oxidation-
659 reduction potential, Environ. Sci. Technol. 45 (2011) 1586–1592.

660 [39] M.J. Truex, V.R. Vermeul, D.P. Mendoza, B.G. Fritz, R.D. Mackley, M. Oostrom, T.W.
661 Wietsma, T.W. Macbeth, Injection of zero-valent iron into an unconfined aquifer using
662 shear-thinning fluids, Ground Water Monit. Remed. 31 (2011) 50–58.

663 [40] M.J. Truex, T.W. Macbeth, V.R. Vermeul, B.G. Fritz, D.P. Mendoza, R.D. Mackley,
664 T.W. Wietsma, G. Sandberg, T. Powell, J. Powers, E. Pitre, M. Michalsen, S.J. Ballock-
665 Dixon, L. Zhong, M. Oostrom, Demonstration of combined zero-valent iron and electrical resistance heating for in situ trichloroethene remediation. Environ. Sci. Technol. (2011) doi: 10.1021/es104266a.

[41] T. Masciangioli, W.X. Zhang, Environmental technologies at the Nanoscale, Environ. Sci. Technol. 37 (2003) 102A–108A.

[42] A. Ghauch, A. Tuqan, H. Abou Assi, Antibiotic removal from water: Elimination of amoxicillin and ampicillin by microscale and nanoscale iron particles, Environ. Pollut. 157 (2009) 1626–1635.

[43] N. Sakulchaicharoen, D.M. O'Carroll, J.E. Herrera, Enhanced stability and dechlorination activity of pre-synthesis stabilized nanoscale FePd particles, J. Contam. Hydrol. 118 (2010) 117–127.

[44] V. Nagpal, A.D. Bokare, R.C. Chikate, C.V. Rode, K.M. Paknikar, Reply to comment on "Reductive dechlorination of γ-hexachlorocyclohexane using Fe–Pd bimetallic nanoparticles", by C. Noubactep, J. Hazard. Mater. (2011) doi:10.1016/j.jhazmat.2011.04.015.

[45] C. Anstice, C. Alonso, F.J. Molina, Cover cracking as a function of bar corrosion: part I- experimental test, Materials and structures 26 (1993) 453–464.

[46] S. Caré, Q.T. Nguyen, V. L'Hostis, Y. Berthaud, Mechanical properties of the rust layer induced by impressed current method in reinforced mortar, Cement Concrete Res. 38 (2008) 1079–1091.

[47] Y. Zhao, H. Ren, H. Dai, W. Jin, Composition and expansion coefficient of rust based on X-ray diffraction and thermal analysis, Corros. Sci. 53 (2011) 1646–1658.

[48] K.D. Grieger, A. Fjordboge, N.B. Hartmann, E. Eriksson, P.L. Bjerg, A. Baun, Environmental benefits and risks of zero-valent iron nanoparticles (nZVI) for in situ remediation: Risk mitigation or trade-off? J. Contam. Hydrol. 118 (2010) 165–183.

690    [49] R.J. Barnes, C. J. van der Gast, O. Riba, L.E. Lehtovirta, J.I. Prosser, P.J. Dobson, I.P.
691         Thompson, The impact of zero-valent iron nanoparticles on a river water bacterial
692         community, J. Hazard. Mater. 184 (2010) 73–80.

693    [50] M. Diao, M. Yao, Use of zero-valent iron nanoparticles in inactivating microbes, Water
694         Res. 43 (2009) 5243–5251.

695    [51] T. Tervonen, I. Linkov, J.R. Figueira, J. Steevens, M. Chappell, M. Merad, Risk-based
696         classification system of nanomaterials, J. Nanopart. Res. 11 (2009) 757–766.

697    [52] C. Noubactep (2007): Processes of contaminant removal in "$Fe^0$–$H_2O$" systems revisited.
698         The importance of co-precipitation, Open Environ. J. 1, 9–13.

699    [53] C. Noubactep A critical review on the mechanism of contaminant removal in $Fe^0$–$H_2O$
700         systems, Environ. Technol. 29 (2008) 909–920.

701    [54] C. Noubactep, The fundamental mechanism of aqueous contaminant removal by metallic
702         iron, Water SA 36 (2010) 663–670.

703    [55] M.I. Litter, M.E. Morgada, J. Bundschuh, Possible treatments for arsenic removal in
704         Latin American waters for human consumption, Environ. Pollut. 158 (2010) 1105–1118.

705    [56] R.A. Crane, M. Dickinson, I.C. Popescu, T.B. Scott, Magnetite and zero-valent iron
706         nanoparticles for the remediation of uranium contaminated environmental water, Water
707         Res. 45  (2011) 2931–2942.

708    [57] O. Celebi, C. Uzum, T. Shahwan, H.N. Erten, A radiotracer study of the adsorption
709         behavior of aqueous $Ba^{2+}$ ions on nanoparticles of zero-valent iron, J. Hazard. Mater.
710         148 (2007) 761–767.

711    [58]  X.Q. Li, W.X. Zhang, Sequestration of metal cations with zerovalent iron
712         nanoparticles—a study with high resolution X-ray photoelectron spectroscopy (HR-
713         XPS), J. Phys. Chem. C 111 (2007) 6939–6946.

714   [59] H.K. Boparai, M. Joseph, D.M. O'Carroll, Kinetics and thermodynamics of cadmium ion
715         removal by adsorption onto nano zerovalent iron particles, J. Hazard. Mater. 186 (2011)
716         458–465.

717   [60] S. Xiao, H. Ma, M. Shen, S. Wang, Q. Huang, X. Shi, Excellent copper(II) removal using
718         zero-valent iron nanoparticle - immobilized hybrid electrospun polymer nanofibrous
719         mats, Colloids Surf. A: Physicochem. Eng. Aspects 381 (2011) 48–54.

720   [61] E.J. Reardon, R. Fagan, J.L. Vogan, A. Przepiora, Anaerobic corrosion reaction kinetics
721         of nanosized iron, Environ. Sci. Technol. 42 (2008) 2420–2425.

722   [62] M. Stratmann, J. Müller, The mechanism of the oxygen reduction on rust-covered metal
723         substrates, Corros. Sci. 36 (1994) 327–359.

724   [63] C. Noubactep, T. Licha, T.B. Scott, M. Fall, M. Sauter, Exploring the influence of
725         operational parameters on the reactivity of elemental iron materials, J. Hazard. Mater.
726         172 (2009) 943–951.

727   [64] S.H. Behrens, D.I. Christl, R. Emmerzael, P. Schurtenberger, M. Borkovec, Charging
728         and aggregation properties of carboxyl latex particles: experiments versus DLVO
729         theory, Langmuir 21 (2000) 2566–2575.

730   [65] M. Dickinson, T.B. Scott, The application of zero-valent iron nanoparticles for the
731         remediation of a uranium-contaminated waste effluent, J. Hazard. Mater. 178 (2010)
732         171–179.

733   [66] R.L. Johnson, R.B. Thoms, R.O.B. Johnson, J. Nurmi, P.G. Tratnyek, Mineral
734         precipitation upgradient from a zero-valent iron permeable reactive barrier, Ground
735         Water Monit. Rem. 28 (2008) 56–64.

736   [67] Y. Wu, J. Zhang, Y. Tong, X. Xu, Chromium (VI) reduction in aqueous solutions by
737         $Fe_3O_4$-stabilized $Fe^0$ nanoparticles, J. Hazard. Mater. 172 (2009) 1640–1645.

738  [68] Z. Fang, X. Qiu, J. Chen, X. Qiu, Degradation of the polybrominated diphenyl ethers by
739       nanoscale zero-valent metallic particles prepared from steel pickling waste liquor,
740       Desalination 267 (2011) 34–41.

741  [69] Z.LvL. Jiang, W. Zhang, Q. Du, B. Pan, L. Yang, Q. Zhang, Nitrate reduction using
742       nanosized zero-valent iron supported by polystyrene resins: Role of surface functional
743       groups. Water Res. 45 (2011) 2191–2198.

744  [70] M. Tong, S. Yuan, H. Long, M. Zheng, L. Wang, J. Chen, Reduction of nitrobenzene in
745       groundwater by iron nanoparticles immobilized in PEG/nylon membrane, J. Contam.
746       Hydrol. 122 (2011) 16–25.

747  [71] S. Yuan, Z. Zheng, X.-Z. Meng, J. Chen, L. Wang,  Surfactant  mediated  HCB
748       dechlorination in contaminated soils and sediments by micro and nanoscale Cu/Fe
749       Particles. Geoderma 159 (2010) 165–173.

750  [72] N. Zhu, H. Luan, S. Yuan, J. Chen, X. Wu, L. Wang, Effective dechlorination of HCB by
751       nanoscale Cu/Fe particles. J. Hazard. Mater. 176 (2010), 1101–1105.

752  [73] B.S. Kadu, Y.D. Sathe, A.B. Ingle, R.C. Chikate, K.R. Patil, C.V. Rode, Efficiency and
753       recycling capability of montmorillonite supported Fe–Ni bimetallic nanocomposites
754       towards hexavalent chromium remediation, Appl. Catal. B: Environ. 104  (2011) 407–
755       414.

756  [74] C. Noubactep, S. Caré, F. Togue-Kamga, A. Schöner, P. Woafo, Extending service life
757       of household water filters by mixing metallic iron with sand, Clean – Soil, Air, Water 38
758       (2010) 951–959.

759  [75] V. Tarvainen, A. Ranta-Maunus, A. Hanhijärvi, H. Forsén, The effect of drying and
760       storage conditions on case hardening of scots pine and norway spruce timber, Maderas.
761       Ciencia y tecnología 8 (2006) 3–14.

762   [76] W.J.N. Fernando, A.L. Ahmad, S.R. Abd. Shukor, Y.H. Lok, A model for constant
763       temperature drying rates of case hardened slices of papaya and garlic, J. Food Eng. 88
764       (2008) 229 –238

765   [77] Y. Zhuang, S. Ahn, A.L. Seyfferth, Y. Masue-Slowey, S. Fendorf, R.G. Luthy,
766       Dehalogenation of polybrominated diphenyl ethers and polychlorinated biphenyl by
767       bimetallic, impregnated, and nanoscale zerovalent iron, Environ. Sci. Technol. 45
768       (2011) 4896–4903.

769   [78] N. Moraci, P.S. Calabrò, Heavy metals removal and hydraulic performance in zero-
770       valent iron/pumice permeable reactive barriers, J. Environ. Manag. 91 (2010) 2336–
771       2341.

772   [79] Y.N. Vodyanitskii, The role of iron in the fixation of heavy metals and metalloids in
773       soils: a review of publications, Eurasian Soil Sci. 43 (2010) 519–532.

774   [80] S.R. Kanel, J.-M. Greneche, H. Choi, Arsenic(V) Removal from groundwater using nano
775       scale zero-valent iron as a colloidal reactive barrier material, Environ. Sci. Technol. 40
776       (2006) 2045–2050.

777   [81] D.D.J. Antia, Modification of aquifer pore-water by static diffusion using nano-zero-
778       valent metals. Water 3 (2011) 79–112.

779   [82] M. Flury, H. Flühler, Brilliant Blue FCF as a dye tracer for solute transport studies. A
780       toxicological review, J. Environ. Qual. 23 (1994) 1108–1112.

781   [83] A.E. Fryar, F.W. Schwartz, Hydraulic-conductivity reduction, reaction-front propagation,
782       and preferential flow within a model reactive barrier, J. Contam. Hydrol. 32 (1998) 333–
783       351.

784   [84] J. Simunek, N.J. Jarvis, M.T. van Genuchten, A. Gardenas, Review and comparison of
785       models for describing non-equilibrium and preferential flow and transport in the vadose
786       zone, J. Hydrol. 272 (2003) 14–35.

[85] B.E. Clothier, S.R. Green, M. Deurer, Preferential flow and transport in soil: progress and prognosis, Eur. J. Soil Sci. 59 (2008) 2–13.

[86] S.E. Allaire, S. Roulier, A.J. Cessna, Quantifying preferential flow in soils: A review of different techniques, J. Hydrol. 378 (2009) 179–204.

[87] D.C. McMurty, R.O. Elton, New approach to in-situ treatment of contaminated groundwaters, Environ. Progr. 4/3 (1985) 168–170.

[88] M.D. Mackay, J.A. Cherry, Groundwater contamination: Pump-and-treat remediation, Environ. Sci. Technol. 23 (1989) 630–636.

[89] R.C. Starr, J.A. Cherry, In situ remediation of contaminated Ground water: The funnel-and-Gate System, Ground Water 32 (1994) 465–476.

[90] M. Min, H. Xu, J. Chen, M. Fayek, Evidence of uranium biomineralization in sandstone-hosted roll-front uranium deposits, northwestern China, Ore Geol. Rev. 26 (2005) 198–206.

[91] M. Sidborn, I. Neretnieks, Long term redox evolution in granitic rocks: Modelling the redox front propagation in the rock matrix, Appl. Geochem. 22 (2007) 2381–2396.

[92] B. Gu, T.J. Phelps, L. Liang, M.J. Dickey, Y. Roh, B.L. Kinsall, A.V. Palumbo, G.K. Jacobs, Biogeochemical dynamics in zerovalent iron columns: implications for permeable reactive barriers, Environ. Sci. Technol. 33 (1999) 2170–2177.

[93] C. Su, R.W. Puls, Arsenate and arsenite removal by zerovalent iron: kinetics, redox transformation, and implications for in situ groundwater remediation, Environ. Sci. Technol. 35 (2001) 4562–4568.

[94] Y. Furukawa, J.-W. Kim, J. Watkins, R.T. Wilkin, Formation of ferrihydrite and associated iron corrosion products in permeable reactive barriers of zerovalent iron, Environ. Sci. Technol. 36 (2002) 5469–5475.

811 [95] T. Kohn, J.T. Kenneth, A. Livi, A.L. Roberts, P.J. Vikesland, Longevity of granular iron

812 in groundwater treatment processes: corrosion product development, Environ. Sci.

813 Technol. 39 (2005) 2867–2879.

814 [96] C.D. Palmer, P.R. Wittbrodt, Processes affecting the remediation of chromium-

815 contaminated sites, Environ. Health Perspect. 92 (1991) 25–40.

816 [97] S. Nesic, Key issues related to modelling of internal corrosion of oil and gas pipelines –

817 A review, Corros. Sci. 49 (2007) 4308–4338.

818 [98] J.R. Kiser, Bruce A. Manning, Reduction and immobilization of chromium(VI) by

819 iron(II)-treated faujasite. J. Hazard. Mater. 174 (2010) 167–174.

820 [99] M.J. Lottering, L. Lorenzen, N.S. Phala, J.T. Smit, G.A.C. Schalkwyk, Mineralogy and

821 uranium leaching response of low grade South African ores, Miner. Eng. 21 (2008) 16–

822 22.

823 [100] Y. Jiao, C. Qiu, L. Huang, K. Wu, H. Ma, S. Chen, L. Ma, L. Wu, Reductive

824 dechlorination of carbon tetrachloride by zero-valent iron and related iron corrosion,

825 Appl. Catal. B: Environ. 91 (2009) 434–440.

826 [101] R.J. Crawford, I.H. Harding, D.E. Mainwaring, Adsorption and coprecipitation of single

827 heavy metal ions onto the hydrated oxides of iron and chromium, Langmuir 9 (1993)

828 3050–3056.

829 [102] R.J. Crawford, I.H. Harding, D.E. Mainwaring, Adsorption and coprecipitation of

830 multiple heavy metal ions onto the hydrated oxides of iron and chromium, Langmuir 9

831 (1993) 3057–3062.

832 [103] K. Eusterhues, T. Rennert, H. Knicker, I. Kgel-Knabner, K.U. Totsche, U.

833 Schwertmann, Fractionation of organic matter due to reaction with ferrihydrite:

834 Coprecipitation versus adsorption, Environ. Sci. Technol. 45 (2011) 527–533.

835     [104] W.P. Johnson, H. Ma, E. Pazmino, Straining credibility: A general comment regarding
836         common arguments used to infer straining as the mechanism of colloid retention in
837         porous media, Environ. Sci. Technol. 45 (2011) 3831–3832.

838     [105] M. Kalin, W.N. Wheeler, G. Meinrath, The removal of uranium from mining waste
839         water using algal/microbial biomass, J. Environ. Radioact. 78 (2005) 151–177.

840     [106] C. Noubactep, Metallic iron for safe drinking water worldwide, Chem. Eng. J. 165
841         (2010) 740–749.

842     [107] D. Pokhrel, T. Viraraghavan, Arsenic removal in an iron oxide-coated fungal biomass
843         column: Analysis of breakthrough curves, Biores. Technol. 99 (2008) 2067–2071.

844     [108] D. Pokhrel, B.S. Bhandari, T. Viraraghavan, Arsenic contamination of groundwater in
845         the Terai region of Nepal: an overview of health concerns and treatment options,
846         Environ. Int. 35 (2009) 157–161.

847     [109] D. Pokhrel, T. Viraraghavan, Biological filtration for removal of arsenic from drinking
848         water, J. Environ. Manage. 90 (2009) 1956–1961.

849     [110] J.S. Morrison, R.R. Sprangler, Chemical barriers for controlling groundwater
850         contamination, Environ. Progr. 12 (1993) 175–181.

851     [111] J.S. Morrison, R.R. Sprangler, S.A. Morris, Subsurface injection of dissolved ferric
852         chloride to form a chemical barrier: Laboratory investigations, Ground Water 34 (1996)
853         75–83.

854     [112] K. Hanna, J.-F. Boily, Sorption of two naphthoic acids to goethite surface under flow
855         through conditions, Environ. Sci. Technol. 44 (2010) 8863–8869.

856     [113] A. Ghauch, H. Abou Assi, S. Bdeir Aqueous removal of diclofenac by plated elemental
857         iron: Bimetallic systems, J. Hazard. Mater. 182 (2010) 64–74.

858     [114] C. Noubactep, Aqueous contaminant removal by metallic iron: Is the paradigm shifting?
859         Water SA 37 (2011) xy–zt.

860  [115] R.L. Reynolds, M.B. Goldhaber, Origin of a south Texas roll-type uranium deposit: I.

861       Alteration of iron-titanium oxide minerals, Econ. Geol. 73 (1978) 1677–1689.

862  [116] J. Posey-Dowty, E. Axtmann, D. Crerar, M. Borcsik, A. Ronk, W. Woods, Dissolution

863       rate of uraninite and uranium roll-front ores. Econ. Geol. 82 (1987) 184–194.

864  [117] L. Romero, I. Neretnieks, L. Moreno, Movement of the redox front at the Osamu

865       Utsumi uranium mine, Poços de Caldas, Brazil, J. Geochem. Explor. 45 (1992) 471–

866       501.

867  [118] D. Read, T.A. Lawless, R.J. Sims, K.R. Butter, Uranium migration through intact

868       sandstone cores, J. Cont. Hydrol., 13 (1993) 277–289.

869

870    **Table 1**: Results of a web-search in 7 selected relevant journals demonstrating the current

871              interest within academia for the nano-$Fe^0$ technology.

872

873

| Journal | Impact Factor | Issues ($year^{-1}$) | Search's results | | |
|---|---|---|---|---|---|
| | | | Period | Total | 2011 |
| Environ. Sci. Technol. | 4.630 | 24 | 1995 to 2011 | 157 | 13 |
| J. Hazard. Mater. | 4.144 | 33 | 2004 to 2011 | 86 | 15 |
| Chemosphere | 3.253 | 44 | 2000 to 2011 | 49 | 5 |
| Water Res. | 4.355 | 20 | 2005 to 2011 | 31 | 9 |
| Chem. Eng. J. | 2.816 | 30 | 2008 to 2011 | 15 | 6 |
| Desalination | 2.034 | 48 | 2008 to 2011 | 12 | 9 |
| Environ. Pollut. | 3.426 | 12 | 2007 to 2011 | 12 | 1 |
| Appl. Catal. B | 5.252 | 32 | 2009 to 2011 | 3 | 1 |
| | | | **Total** | **365** | **59** |

874

875 **Table 2:** Relevant redox couples for the process of aqueous $Fe^0$ dissolution and oxide scale

876 formation in a passive remediation $Fe^0/H_2O$ system. These processes are thermodynamically

877 the same for all $Fe^0$ particle sizes. Observed differences are due to kinetics aspects.

878

| Electrode reactions | | | Eq. |
|---|---|---|---|
| $Fe^0$ | $\Leftrightarrow$ | $Fe^{2+} + 2\ e^-$ | (1) |
| Oxic conditions | | | |
| $O_2 + 2\ H_2O + 4\ e^-$ | $\Leftrightarrow$ | $4\ OH^-$ | (2a) |
| $2\ H_2O + 2\ e\text{-}$ | $\Leftrightarrow$ | $H_2 + 2\ OH^-$ | (2b) |
| Anoxic conditions | | | |
| $O_2 + 4\ H^+ + 4\ e^-$ | $\Leftrightarrow$ | $2\ H_2O$ | (3a) |
| $2\ H^+ + 2\ e^-$ | $\Leftrightarrow$ | $H_2$ | (3b) |

879

**Table 3**: Summary of the values of the number of particles contained in 1 kg of each material, the number of layer making up each particle and estimation of the relative time ($\tau$). The life span of nano-Fe$^0$ is operationally considered as the unit of time while assuming uniform corrosion. $\tau$ coincides with the ratio of the number of layers of Fe in each particle to that of nano-Fe$^0$. The ratio of the number of particles in individual systems is also given.

| Size | d | $n_{particles}$ | $n_{layers}$ | $n_{layers}/n_{nano}$ | $n_{nano}/n_{particles}$ | $\tau$ |
|------|------|------|------|------|------|------|
| | ($\mu$m) | (-) | (-) | (-) | (-) | (-) |
| **Nano-Fe$^0$** | $25*10^{-3}$ | $1.96*10^{18}$ | 87.2 | 1.0 | 1.0 | 1.0 |
| **$\mu$m-Fe$^0$** | 25 | $1.96*10^9$ | $87.2*10^3$ | $10^3$ | $10^9$ | $10^3$ |
| **mm-Fe$^0$** | 250 | $1.96*10^6$ | $87.2*10^4$ | $10^4$ | $10^{12}$ | $10^4$ |
| **mm-Fe$^0$** | 1000 | $3.06*10^4$ | $3.49*10^5$ | $4*10^4$ | $6.4*10^{13}$ | $4*10^4$ |

**Table 4**: Estimation of the value of the life span ($t_\infty$) of a nano-$Fe^0$ particle with 25 nm diameter for barrier life spans (t) from 10 to 40 years. The considered conventional reactive wall contains granular $Fe^0$ with a diameter of 1 mm. For comparison the relative life span (in days and years) of the micrometric particles is given.

| t | (years) | 5 | 10 | 15 | 20 | 25 | 30 | 35 | 40 |
|---|---|---|---|---|---|---|---|---|---|
| $t_{\mu m}$ | (days) | 45.7 | 91.3 | 137.0 | 182.6 | 228.3 | 273.9 | 319.6 | 365.3 |
| $t_{\mu m}$ | (years) | 0.2 | 0.3 | 0.5 | 0.7 | 0.9 | 1.0 | 1.2 | 1.4 |
| $t_\infty$ | (hours) | 1.1 | 2.2 | 3.3 | 4.4 | 5.5 | 6.6 | 7.7 | 8.8 |

**Table 5**: Estimation of the extent of porosity loss ($\Delta V$) due to the volumetric expansion of iron corrosion for $Fe^0$ particles of different sizes. The operational unit of time is arbitrarily the time to nano-$Fe^0$ depletion ($t_\infty$). $V_0$ is the volume occupied by the initial $Fe^0$ particles; $V_\infty$ is the volume occupied by residual $Fe^0$ and in-situ formed corrosion products. $\Delta V$ corresponds to the volume of pore occupied by the volumetric expansion of corrosion products. $m_{consumed}$ = mass of $Fe^0$ consumed; $\nu$ = percent of $Fe^0$ depletion, $n_{Fe(II)}$ = number of moles of corroded $Fe^0$; $n_{Fe3O4}$ = number of moles of generated iron corrosion products, $n_{electrons} = 2*n_{Fe(II)}$ = number of electrons released by corroded iron; and $n_{nano}/n_{electrons}$ is the ratio of the number of electrons produced in by nano-$Fe^0$ to $n_{electrons}$.

| Size | $m_{consumed}$ (kg) | $\nu$ (%) | $V_\infty$ (mL) | $\Delta V$ (mL) | $\Delta V$ (%) | $n_{Fe(II)}$ (moles) | $n_{Fe3O4}$ (moles) | $n_{electrons}$ (moles) | $n_{nano}/n_{electrons}$ (-) |
|---|---|---|---|---|---|---|---|---|---|
| nm-$Fe^0$ | $10^0$ | 100.00 | 264.16 | 137.16 | 108.00 | 17.857 | 5.9524 | 35.714 | 1 |
| µm-$Fe^0$ | $3*10^{-3}$ | 0.30 | 127.41 | 0.41 | 0.32 | 0.053 | 0.0178 | 0.107 | 335 |
| mm-$Fe^0$ | $3*10^{-4}$ | 0.03 | 127.04 | 0.04 | 0.03 | 0.005 | 0.0018 | 0.011 | 3342 |
| mm-$Fe^0$ | $7.5*10^{-5}$ | 0.01 | 127.01 | 0.01 | 0.01 | 0.001 | 0.0004 | 0.003 | 13369 |

909

910

911

912 **Table 6**: Summary of the values of the number of particles contained in 1 kg of each nano-
913       $Fe^0$, the number of layer making up each particle and estimation of the relative time
914       ($\tau$). The life span of the material with the smallest particle site (d = 10 nm) is
915       operationally considered as the unit of time while assuming uniform corrosion. The
916       ratio of the number of particles in 1 kg of d = 10 nm to that of other d values is also
917       given.

918

| d | $n_{particles}$ | $n_{layers}$ | $n_{layers}/n_{10}$ | $n_{10}/n_{particles}$ | $\tau$ |
|---|---|---|---|---|---|
| (nm) | (-) | (-) | (-) | (-) | (-) |
| 10 | $3.06*10^{19}$ | 34.9 | 1.0 | 1.0 | 1.0 |
| 25 | $1.96*10^{18}$ | 87.2 | 2.5 | 16 | 2.5 |
| 50 | $2.45*10^{11}$ | 174.5 | 5.0 | 125 | 5.0 |
| 75 | $7.25*10^{16}$ | 261.7 | 7.5 | 422 | 7.5 |
| 100 | $3.06*10^{16}$ | 348.9 | 10.0 | 1000 | 10.0 |

919

920

 **Table 7**: Summary of the extent of porosity loss ($\Delta V_{pore}$ in %) as 1 kg of nano-$Fe^0$ ($V_0 = 127$

mL) is corroded to various iron oxides. $V_\infty$ is the volume of iron oxide at $Fe^0$

exhaustion. The initial volume of pore ($V_{pore}$) is 530.4 mL and $V'_{pore}$ is the residual

pore volume at $Fe^0$ exhaustion. The absolute value of negative values for $V'_{pore}$

corresponds to the mass of nano-$Fe^0$ which can not oxidize because of lack of space for volumetric expansion.

| Fe species | $\eta$ | $V_\infty$ | $\Delta V$ | $V'_{pore}$ | $\Delta V_{pore}$ |
|---|---|---|---|---|---|
| | (-) | (mL) | (mL) | (mL) | (%) |
| $Fe^0$ | 1 | 127 | 0 | 403.4 | 23.9 |
| $Fe_3O_4$ | 2.08 | 264.2 | 137.2 | 266.2 | 49.8 |
| $Fe_2O_3$ | 2.12 | 269.2 | 142.3 | 261.1 | 50.8 |
| $\alpha$-FeOOH | 2.91 | 369.6 | 242.6 | 160.8 | 69.7 |
| $\gamma$-FeOOH | 3.03 | 384.8 | 258.0 | 145.6 | 72.6 |
| $\beta$-FeOOH | 3.48 | 442.0 | 315.0 | 88.4 | 83.3 |
| $Fe(OH)_2$ | 3.75 | 476.3 | 349.3 | 54.1 | 89.8 |
| $Fe(OH)_3$ | 4.2 | 533.4 | 406.4 | -3.0 | 100.0 |
| $Fe(OH)_3.3H_2O$ | 6.4 | 812.8 | 685.8 | -282.4 | 100.0 |

931     **Figure 1**

932

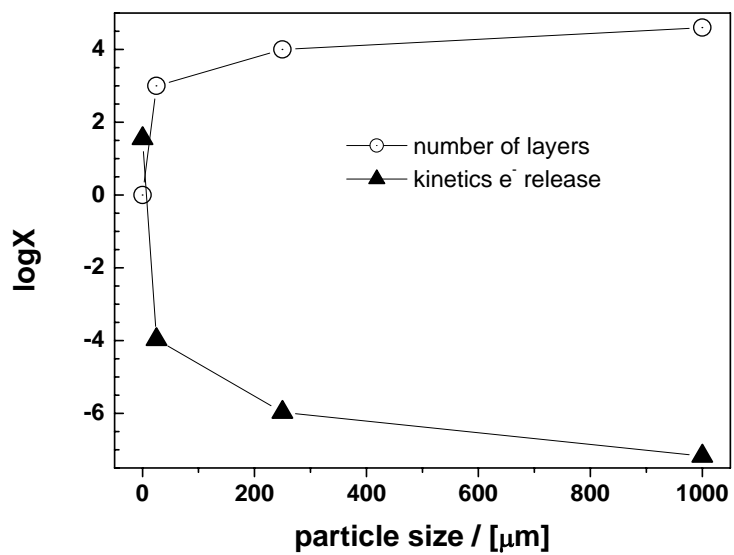

933

934    **Figure 2**

935

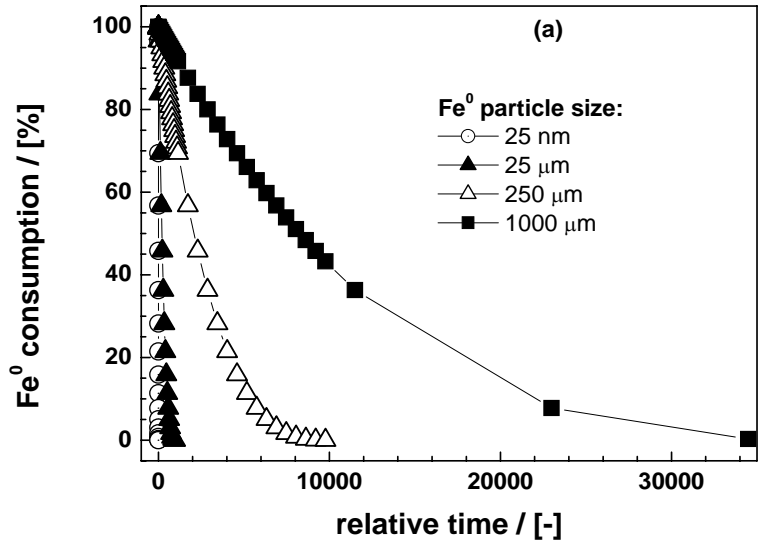

936

937

938

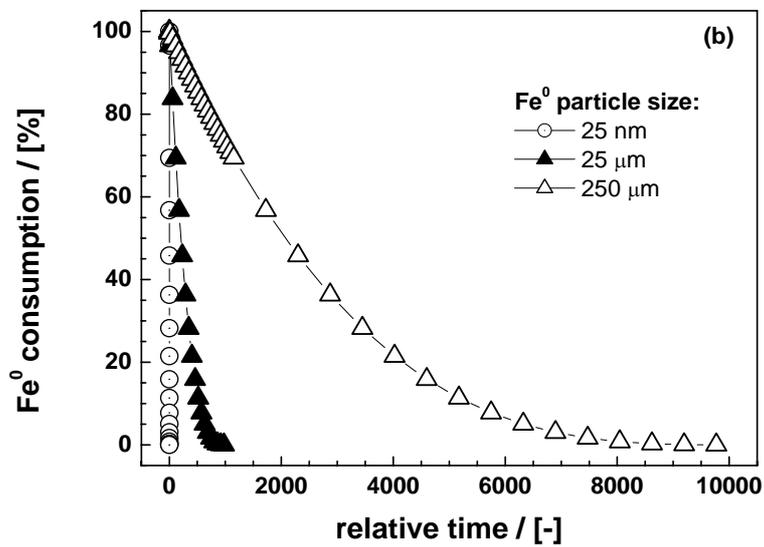

939

940

     **Figure 3**

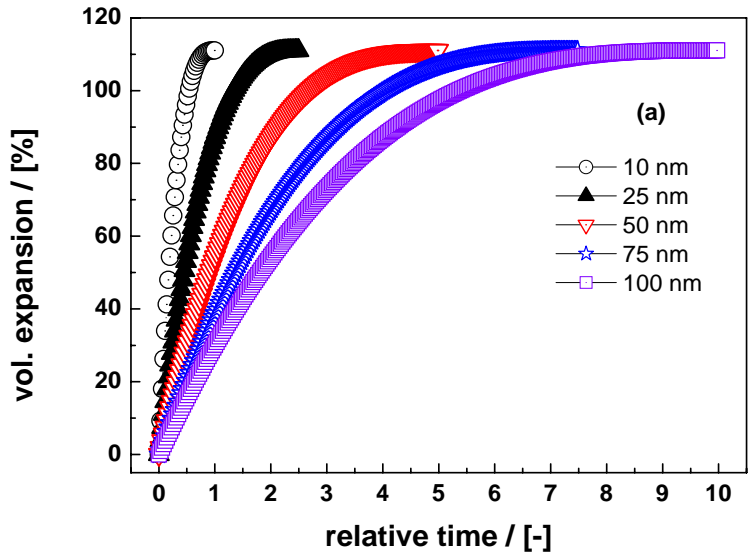

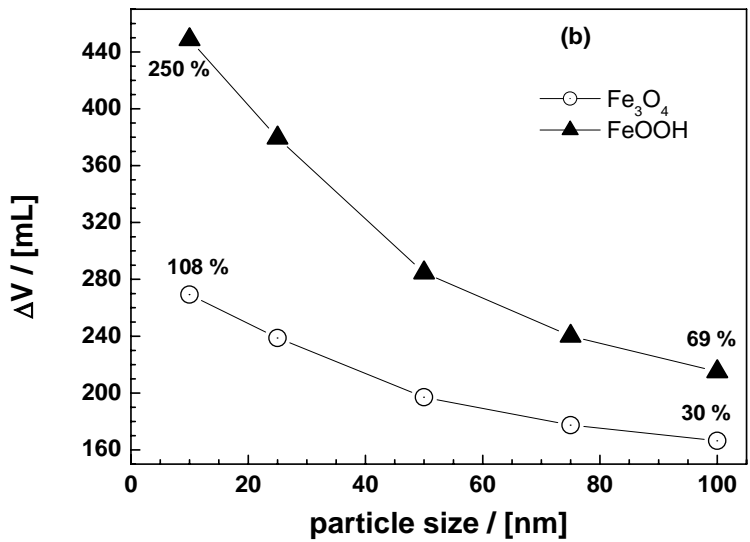

947     **Figure 4**:

948

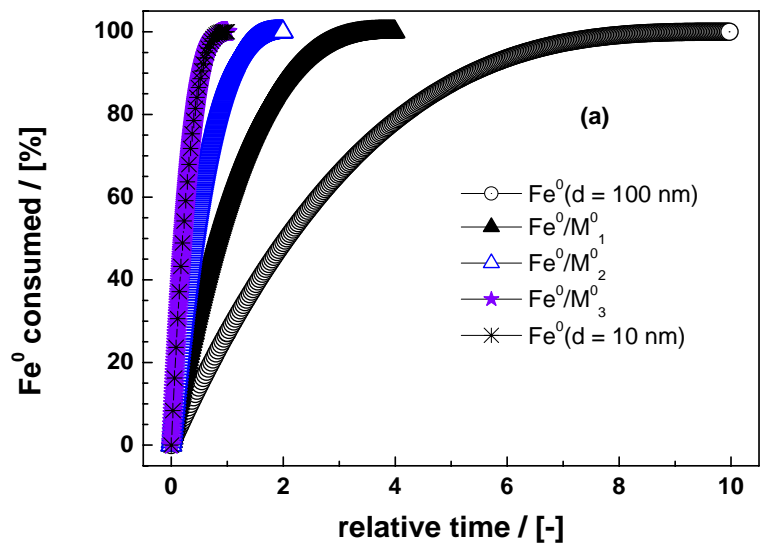

949

950

951     **Figure 5**

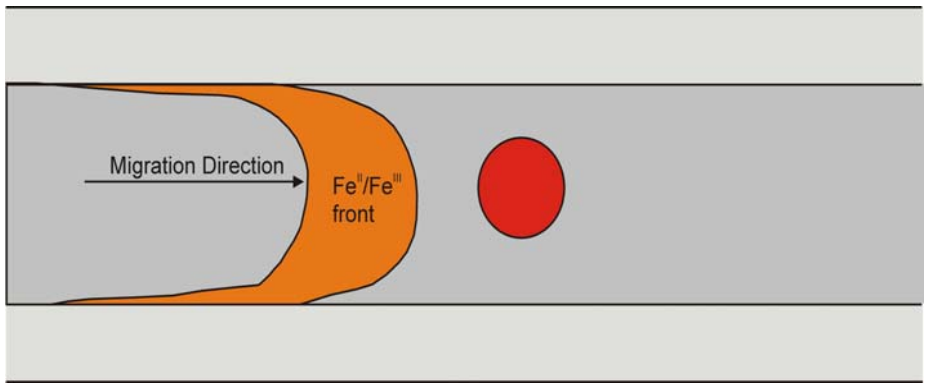

952

953

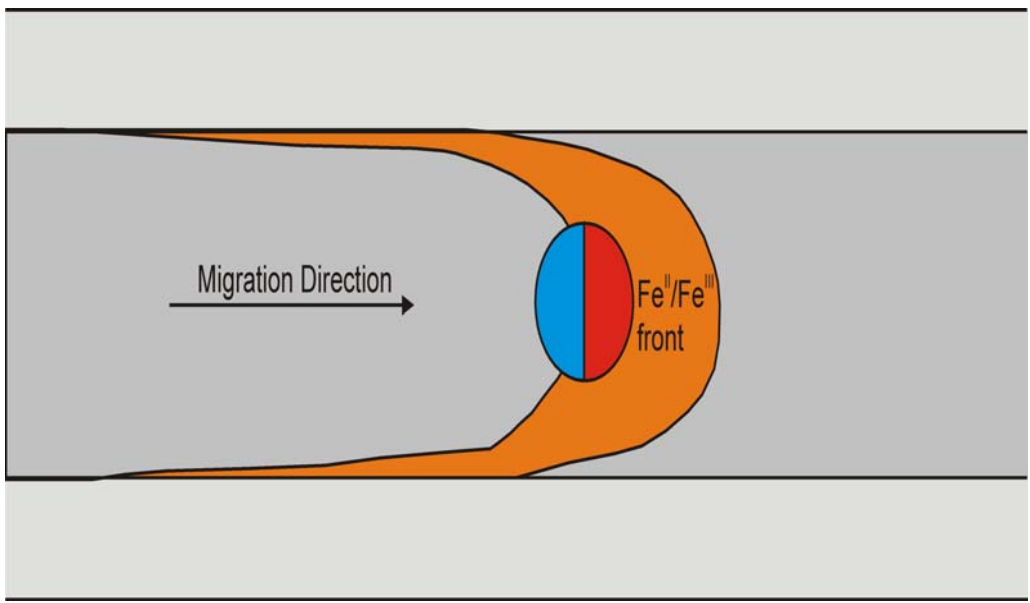

954

955

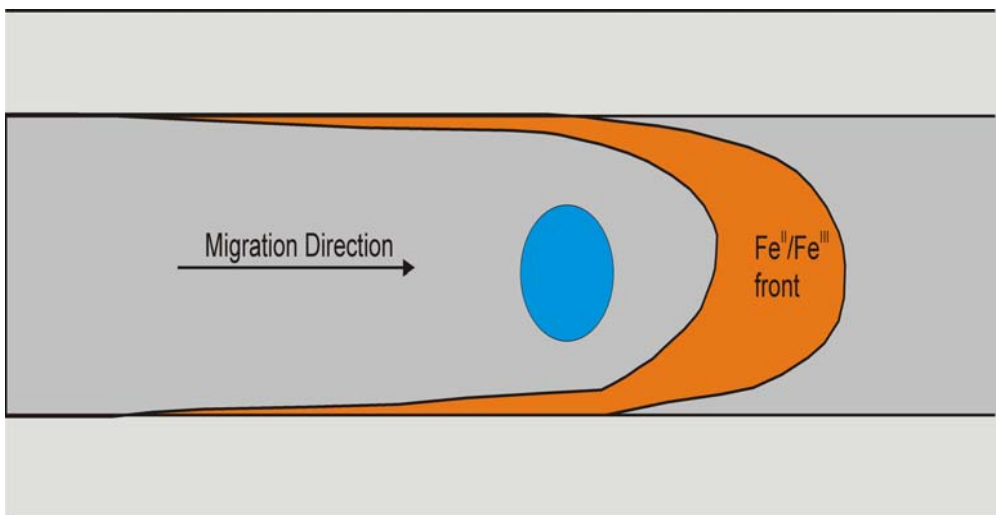

956

957

**Figure captions**

**Figure 1**: Comparison of the evolution of the kinetics of electron release and the number of layers in each $Fe^0$ particle as a function of the particle size. It is shown that smaller particles release huge amounts of electrons within a very short time. Calculations are made for 1 kg of $Fe^0$ material.

**Figure 2**: Kinetics of the process of $Fe^0$ exhaustion at nano-, micro- and millimetre scale as for: (a) d ≤ 1000 μm, and (b) d ≤ 200 μm.

**Figure 3:** Kinetics of the process of porosity loss at nano-scale as characterized by: (a) the percent volumetric expansion for the five considered particle sizes, and (b) the absolute value of ΔV (mL) at $\tau = 1$ for two different iron corrosion products ($Fe_3O_4$ and FeOOH).

**Figure 4:** Calculated extent of $Fe^0$ exhaustion as a function of the relative time ($\tau$) for three ideal bimetallic systems based on the material with 100 nm diameter. The material with 10 nm diameter is represented for comparison.

**Figure 5:** Schematic diagram of the flow process of the U-shaped redox-front through a contaminated zone. Despite the relative importance of preferential flow paths, contaminants are "swept" by the roll-front.